\RequirePackage{lineno}
\documentclass[prb,amsmath,amssymb,floatfix,citeautoscript,preprint]{revtex4}
\usepackage{bm}
\usepackage{epsfig}
\usepackage{subfigure}
\usepackage[usenames]{color}
\usepackage{lineno}
\usepackage{blindtext}

\begin{document}

\title{Topological superconductivity in monolayer transition metal dichalcogenides}
\author{Yi-Ting Hsu$^1$, Abolhassan Vaezi$^2$, Mark H. Fischer$^{3}$, Eun-Ah Kim$^1$}
\affiliation{$^1$Department of Physics, Cornell University, Ithaca, New York 14853, USA \\
Department of Physics, Stanford University, Stanford, California 94305-4060, USA\\$^3$Department of Condensed Matter Physics, Weizmann Institute of Science, Rehovot 7610001, Israel}

\maketitle

\textbf{
Theoretically it has been known that breaking spin-degeneracy and effectively realizing 
“spinless fermions” is a promising path to topological superconductors. 
Yet, topological superconductors are rare to date.
Here, we propose to realize spinless fermions by splitting the spin-degeneracy in momentum space. 
Specifically, we identify monolayer hole-doped transition metal dichalcogenide (TMD)s as candidates for topological superconductors out of such momentum-space-split spinless fermions. 
Although electron-doped TMDs have recently been found superconducting, the observed superconductivity is unlikely topological due to the near spin-degeneracy. 
Meanwhile, hole-doped TMDs with momentum-space-split spinless fermions remain unexplored.
Employing a renormalization group analysis, we propose that the unusual spin-valley locking in hole-doped TMDs together with repulsive interactions selectively favors two topological superconducting states: inter-pocket paired state with Chern number 2 and intra-pocket paired state with finite pair-momentum.
A confirmation of our predictions will open up possibilities for manipulating topological superconductors on the device friendly platform of monolayer TMDs.
}

\section{Introduction}
The quest for material realizations of topological chiral superconductors with nontrivial Chern numbers\cite{ReadGreen,ChiralSCreviewKallin,RMPUnconvScSigrist,RMPtoposcZhang} 
 is fueled by  predictions of exotic signatures, such as Majorana zero modes and quantized Hall effects.
Unfortunately, natural occurrence of bulk topological superconductors are rare with the best candidates being 
superfluid $\rm{^3He}$\cite{HeliumDroplet} and $\rm{Sr_2RuO_4}$\cite{SROreview}. Instead, much recent experimental progress relied on proximity inducing pairing to a spin-orbit-coupled band structure building on the proposal of Fu and Kane\cite{FuKane}. 
Their key insight was that a paired state of
 spinless fermions is bound to be topological and that the surface states of topological insulators are spinless in that the spin-degeneracy is split
in position-space ({$\textbf r$}-space): the two degenerate Dirac surface states with opposite spin textures are spatially separated. Nevertheless despite much experimental progress along this direction
\cite{SuperITIribbonProx,SuperITIprox,JosephsonTIprox,ARPESstmTIscProx,JosephsonTIsurfProx,MajoHintTIproxFisher,MajoSTMTIprox}, 
the confinement of the helical paired state to the interface of the topological insulator and a superconductor limits experimental access to its potentially exotic properties.

Another type of exotic paired states that desires material realization is the finite-momentum-paired states, which has long been pursued since the first proposals by Fulde and Ferrell\cite{FF} and by Larkin and Ovchinnikov\cite{LO}. Most efforts towards realization of such modulated superconductors\cite{FFLOspinImbaColdgas,3DLOZeemanSOCgas}, however, relied on generating finite-momentum pairing using spin-imbalance under an (effective) magnetic field in close keeping with the original proposals. 
Exceptions to such a spin-imbalance-based approach are Ref. ~\onlinecite{FFLOrashbascAgterberg} and  ~\onlinecite{WeylFFLOCho} that made use of spinless Fermi surfaces with shifted centers.
More recently, there have been proposals suggesting modulated paired states in cuprate high-$T_c$ superconductors\cite{PhysRevLett.99.127003,PDWLee,SJTMpdwDavis}. 
However, unambiguous experimental detection of a purely modulated paired state in a solid-state system is lacking.

We note an alternative strategy that could lead to pairing possibilities for both topological and modulated superconductivity: 
 to split the spin-degeneracy of fermions in momentum-space ({$\textbf k$}-space). This approach is essentially dual to the proposal of Fu and Kane and it can be realized in a time-reversal-invariant non-centrosymmetric system when a pair of Fermi surfaces centered at opposite momenta $\pm \textbf k_0$ consist of oppositely spin-polarized electrons [see Fig.~1(a)]. 
When such a spin-valley-locked band structure is endowed with repulsive interactions, conventional pairing will be suppressed. Instead, there will be two distinct pairing possibilities: inter-pocket and intra-pocket pairings, where the latter will be spatially modulated with pairs carrying finite center-of-mass momentum $\pm 2 \textbf k_0$.

What is critical to the success of this strategy is the materialization of  such {$\textbf k$}-space-split spinless fermions.
A new opportunity has arisen with the discovery of a family of
 superconducting two-dimensional (2D) materials, monolayer group-VI transition metal dichalcogenides (TMDs) $MX_2$ ($M=$ Mo, W, $X=$ S, Se)\cite{ExpScNdopedMoS2,ExpScWS2,ExpTMDscIwasa,PressureSCMoS2Iwasa}.  
Although the transition metal atom $M$ and the chalcogen atom $X$ form a 2D hexagonal lattice within a layer as in graphene, monolayer
TMDs differ from graphene in two important ways. Firstly, TMD monolayers are non-centrosymmetric, i.e., inversion symmetry is broken [see Fig.~\ref{TMD}(b) and (c)]. As a result, monolayer TMDs are direct-gap semiconductors\cite{STSMoSe2Gap} with a type of Dresselhaus spin-orbit coupling\cite{spinARPESbulkMoS2,TMD2012} referred to as Ising spin-orbit coupling\cite{Exp_IsingscMoS2_LawYe}. This spin-orbit-coupled band structure leads to the valley Hall effect\cite{TMD2012,VHmonoMoS2McEuen}, which has established TMDs  as experimental platforms for pursuing valleytronics applications\cite{TMD2012,ValleyControlmMoS22012,ValleyOpticmMoS22012,ValleyOpticmWSe22013,VHmonoMoS2McEuen,ValleySpinSwitchReza}.
Our focus, however, is the fact that there is a sizable range of chemical potential in the valence band that could materialize the {$\textbf k$}-space spin-split band structure we desire [see Fig.~\ref{TMD}(d)].
 Secondly, the carriers in TMDs have strong $d$-orbital character and hence, correlation effects are expected to be important. Interestingly,
both intrinsic and pressure-induced superconductivity has been reported in electron-doped (n-doped) TMDs  \cite{ExpScNdopedMoS2,ExpScWS2,ExpTMDscIwasa,PressureSCMoS2Iwasa} with  
the debate regarding the nature of the observed superconducting states still on-going\cite{ndopedMoS2Guinea,KTLaw,fRGndopedMoS2,DFTMRDFSCdomeMoS2,DFTSCCDWMoS2}.

Here, we propose to obtain {$\textbf k$}-space-split spinless fermions by lightly hole-doping (p-doping) monolayer TMDs such that the chemical potential lies between the two spin-split valence bands. We investigate the possible paired states that can be driven by repulsive interactions\cite{KL} in such lightly p-doped TMDs using a perturbative renormalization group (RG) analysis going beyond mean-field theory\cite{KTLaw,pdopedTMDAji}.
We find two distinct topological paired states to be the dominant pairing channels: an inter-pocket chiral (p/d)-wave paired state with Chern number $|C|=2$ and an intra-pocket chiral p-wave paired state with a spatial modulation in phase. The degeneracy can be split by the trigonal warping or Zeeman effect.

\section{Results}
\begin{figure}[!hp]
\begin{centering}
\includegraphics[width=0.8\linewidth]{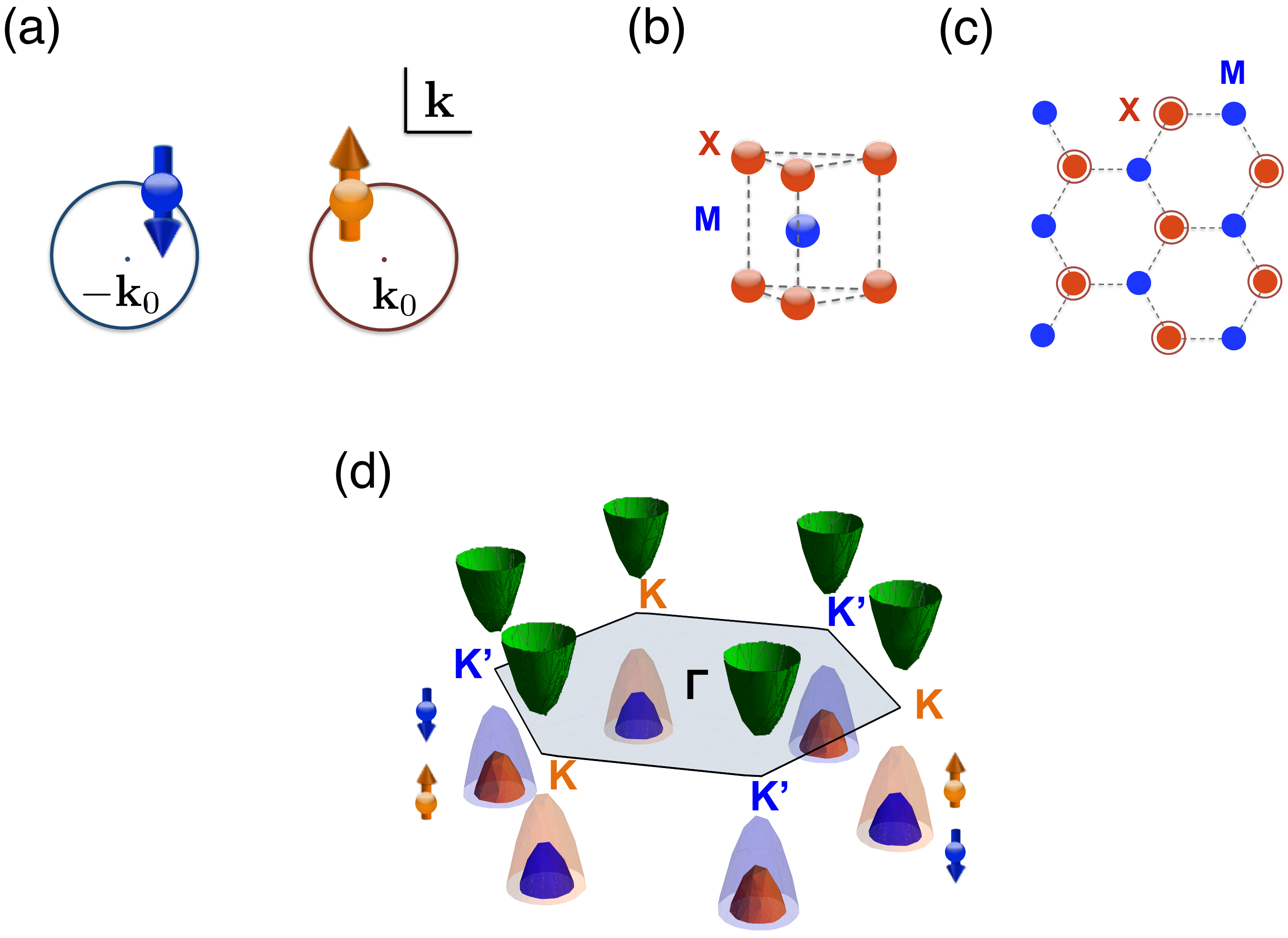}
\end{centering}	
\caption{\textbf{${\textbf k}$-space spin-split in the spin-valley-locked band structure of group IV monolayer TMDs.} (a) Schematic Fermi surface hosting ${\textbf k}$-space-split spinless fermions. 
Here, the two pockets centered at some opposite crystal momenta $\textbf{k}=\pm\textbf{k}_0$ host oppositely spin-polarized electrons (represented by the orange and blue arrows) in a time-reversal-symmetric fashion. 
(b) A sketch for a unit cell of a monolayer TMD. The blue and red spheres represent the transition-metal $M$ atoms and the chalcogen atoms $X$ respectively. (c) A sketch for the top view of the buckled honeycomb lattice of a monolayer TMD. The blue circles represent the transition-metal $M$ atoms and the solid (hollow) red circles represent the chalcogen atoms $X$ above (below) the plane of transition-metal atoms. (d) Schematic low-energy dispersion of a monolayer TMD. The hexagon represents the first Brillouin zone.  
The green paraboloids represent the nearly spin-degenerate conduction band, and the orange and blue paraboloids represent the spin-split valence bands for the spin-up and -down electrons respectively. 
This dispersion is time-reversal symmetric since the spin-splits are opposite near the two valleys $K$ and $K'$ which centered at opposite momenta $\pm\textbf{K}$ with respect to the $\Gamma$ point.}
\label{TMD}
\end{figure}

\textbf{Spin-valley locked fermions in lightly p-doped monolayer TMDs}\\
The generic electronic structure of group IV monolayer TMDs is shown in Fig.~\ref{TMD}(d). The system lacks inversion symmetry [see Fig.~\ref{TMD}(b) and (c)], which leads to a gapped spectrum and a $S_z$-preserving spin-orbit coupling. 
Such Ising spin-orbit coupling\cite{Exp_IsingscMoS2_LawYe} acts as opposite Zeeman fields on the two valleys that  preserves time-reversal symmetry. 
Furthermore the spin-orbit coupling is orbital-selective\cite{SpinsplitMonoTMDs2011} and selectively affects the valence band with a large spin-split 
\cite{spinARPESbulkMoS2}.

\begin{figure}[!hp]
\begin{centering}
\includegraphics[width=1.0\linewidth]{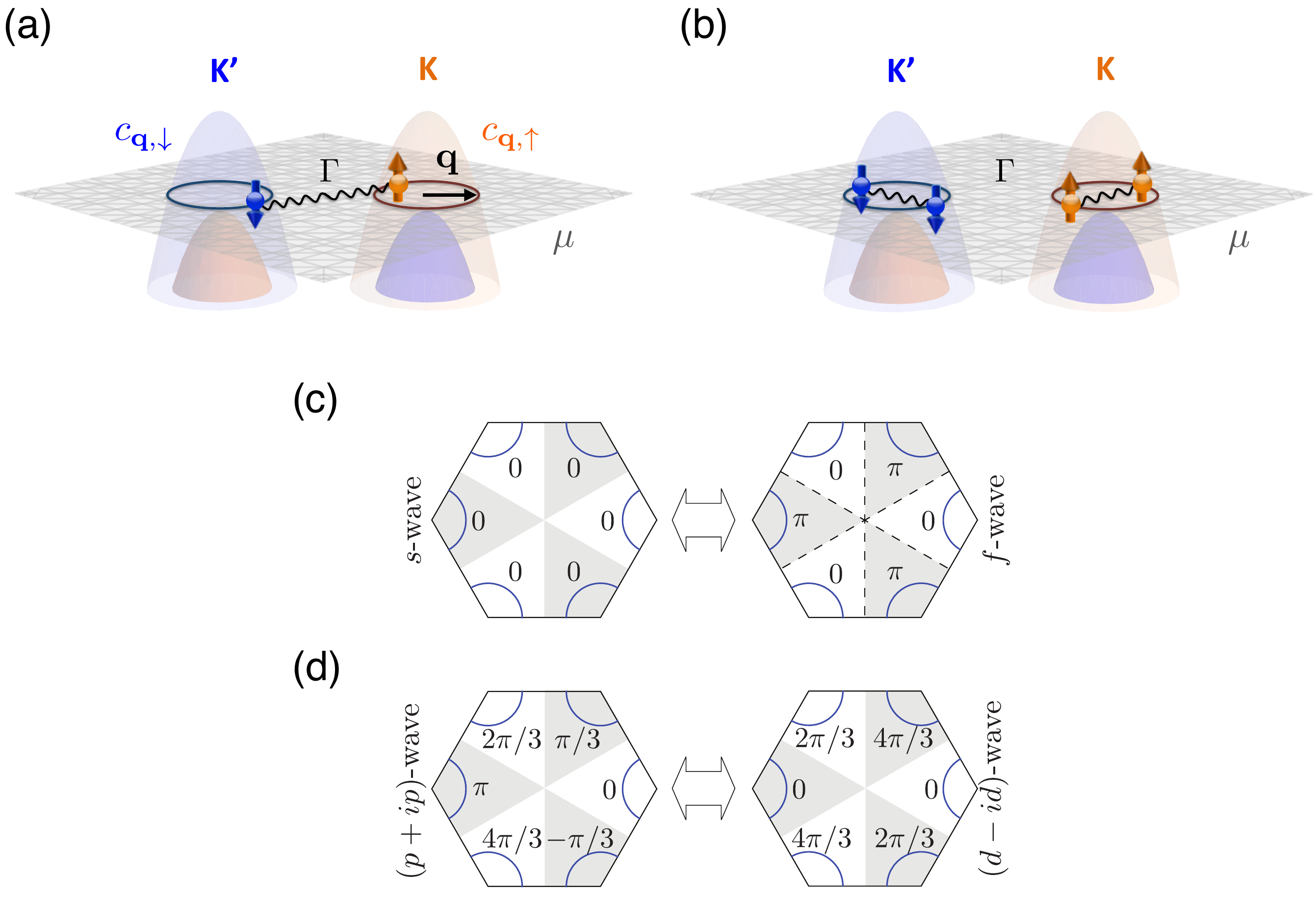}
\end{centering}
\caption{\textbf{Symmetry-distinct pairing channels in a lightly p-doped monolayer TMD.} 
The two oppositely spin-polarized Fermi surfaces centered at $K$ and $K'$ valleys (represented by the maroon and blue circles) can develop (a) inter-pocket pairing or (b) intra-pocket pairing. 
Here, $c_{\textbf{q},\uparrow}$ ($c_{\textbf{q},\downarrow}$) denotes the annihilation operator for spin-up (-down) electrons on the pocket at valley $K$ ($K'$), and $\textbf{q}$ denotes the momentum relative to the pocket centers. 
(c) and (d) are candidate gap functions for inter-pocket pairing allowed by the point group $C_{3v}$. 
Each hexagon represents the first Brillouin zone where the curves around the corners within the unshaded (shaded) wedges are segments of Fermi surfaces around valley $K$ ($K'$).  
Due to the broken $C_6$ rotations (expressed by the shaded wedges), the gap structures of (c) $s$-wave and $f$-wave both belong to the same irreducible representation $A_1$ and can thus mix. 
Similarly, the gap structures of (d) $p$-wave and $d$-wave both belong to the two-dimensional irreducible representation $E$ and can mix as well. 
The number in each wedge labels the angle 
corresponding to the phase 
of each gap function at the midpoint of the Fermi surface segment in the wedge.
Note that the $(p+ip)$- and $(d-id)$-waves have the same phase-winding pattern on each pocket around respective valley centers.}

\label{pairing}
\end{figure}

By lightly p-doping the TMDs with the chemical potential $\mu$ {\textit between} the spin-split valence bands, spin-valley locked fermions can be achieved near the two valleys [see Fig.~\ref{pairing}(a) and (b)]. 
Assuming negligible trigonal warping at low doping, 
we can use a single label $\tau=\uparrow,\downarrow$ to denote the valley and the spin.
Denoting the momentum measured from appropriate valley centers $\pm \textbf{K}$ by $\textbf{q}$, the kinetic part of the Hamiltonian density is 

\begin{align}
H^p_{0}&=\sum_{\textbf{q},\tau} \left(-\frac{q^2}{2m}-\mu\right)c^{\dagger}_{\textbf{q},\tau}c_{\textbf{q},\tau},
\label{eq:H0p}
\end{align}
where $\mu$ is the chemical potential, $m$ is the effective mass of the valence band, and $c_{\textbf{q},\uparrow}\equiv \psi_{\textbf{K}+\textbf{q},\uparrow}$ and $c_{\textbf{q},\downarrow}\equiv \psi_{-\textbf{K}+\textbf{q},\downarrow}$
each annihilates a spin-up electron with momentum $\textbf{q}$ relative to the valley center $\textbf{K}$ or a spin-down electron with momentum $\textbf{q}$ relative to the valley center $-\textbf{K}$ [see Fig.~\ref{pairing}(a)]. 
Hence, the spin-valley locked two-valley problem is now mapped to a problem with a single spin-degenerate Fermi pocket. Nonetheless, the possible paired states with total spin $\tau_z=\pm1$ and $\tau_z=0$ in fact represent the novel possibilities of intra-pocket modulated pairings with total $\tau_z=\pm1$ and inter-pocket pairing with total $\tau_z=0$ respectively
[see Fig.~\ref{pairing}(a) and (b)]. 
 
\textbf{Pairing possibilities}\\ 
To discuss the pairing symmetries of the two pairing possibilities, it is convenient to define the partial-wave channels $\tilde{l}$ with respect to the two valley centers $\pm \textbf{K}$.
Since a total spin $\tau_z=\pm1$ intra-pocket pair consists two electrons with equal spin, Pauli principle dictates such pairing to be in a state with odd partial wave $\tilde{l}$. Stepping back to microscopics, such pairs carry finite center-of-mass momentum $\pm2\textbf{K}$ and form two copies of phase-modulated superconductor\cite{FF}. This case may or may not break time-reversal symmetry 
due to the absence of locking between the $\tilde{l}$s of the two pockets $\tau=\uparrow,\downarrow$.
For the total $\tau_z=0$ inter-pocket pairing, 
the allowed symmetries of a superconducting state is further restricted by the underlying $C_{3v}$ 
symmetry of the lattice. In particular, the absence of an inversion center allows the pairing wavefunction in each irreducible representation to be a mixture between parity-even and -odd functions with respect to the $\Gamma$ point\cite{SrPtAsGapClass}. Specifically, $s$-wave mixes with $f$-wave and $d$-wave mixes with $p$-wave [see Fig.~\ref{pairing}(c) and (d)]. 
Among the irreducible representations of   
$C_{3v}$,  
two fully gapped possibilities are the trivial $A_1$ representation which amounts to $(s/f)$-wave pairing  ($\tilde{l}=0$) and a chiral superposition of the two-dimensional $E$ representation which amounts to a mixture of $p\pm ip$ and $d\mp id$ pairing ($|\tilde{l}|=1$).  The mixing implies that the non-topological $f$-wave channel that is typically dominant in trigonal systems as a way of avoiding repulsive interaction will be blocked together with $s$-wave by the repulsive interaction in the p-doped TMDs. Hence it is clear that the pairing instability in $|\tilde{l}|=1$ channel is all one needs for topological pairing in the p-doped TMDs. 

\textbf{Two distinct topological paired states}\\ 
To investigate the effects of the repulsive interactions between transition metal $d$-orbitals, 
we take the microscopic interaction to be the Hubbard interaction, 
which is the most widely studied pardignamtic model of strongly correlated electronic systems

\begin{equation}
H'(W)=\sum_{i}Un_{i,\uparrow}n_{i,\downarrow},
\label{eq:Hbare}
\end{equation}
where $W$ is the ultra-violet energy scale, $U>0$, and $n_{i,s}$ is the density of electrons with spin $s$ on site $i$. By now it is well-established that the interaction that is purely repulsive at the microscopic level can be attractive in anisotropic channels for low energy degrees of freedom, i.e., fermions near Fermi surface. The perturbative RG approach has been widely used to demonstrate this principle on various correlated superconductors. 
For the model of p-doped TMDs defined by Eqs.~\eqref{eq:H0p} and \eqref{eq:Hbare}, the symmetry-allowed effective interactions at an intermediate energy scale $\Lambda_0\gtrsim 0$ close to the Fermi level in the Cooper channel (see Supplementary Note 1) would be: 

\begin{equation}
H_{\rm{eff}}'(\Lambda_0)=\sum_{\textbf{q},\textbf{q}',\tau,\tau'}
g_{\tau,\tau'}^{(0)}(\textbf{q},\textbf{q}')c^{\dagger}_{\textbf{q}',\tau}c^{\dagger}_{-\textbf{q}',\tau'}c_{-\textbf{q},\tau'}c_{\textbf{q},\tau},
\label{eq:geff}
\end{equation}
where $\textbf{q}$ and $\textbf{q}'$ are the incoming and outgoing momenta. 
Now, the remaining task is to derive the effective inter- and intra-pocket interactions $g_{\uparrow,\downarrow}(\textbf{q},\textbf{q}')$ and $g_{\uparrow,\uparrow}(\textbf{q},\textbf{q}')$ perturbatively in the microscopic repulsion $U$ and check to see if attraction occur in the $|\tilde{l}|=1$ channel (see Methods and Supplementary Note 2). 

Before going into the details of calculation, it is important to note that isotropic pairing with $\tilde{l}=0$ is forbidden by Pauli principle in the total $\tau_z=\pm1$ channel and blocked by the bare repulsive interaction in the total $\tau_z=0$ channel. Hence we need to look for attraction in the  anisotropic $\tilde{l}\neq 0$ channel, which is given by the momentum-dependent part of $g^{(0)}_{\tau\tau'}$. With our assumption of isotropic dispersion at low-doping, one needs to go to the two-loop order to find momentum dependence in the effective interaction. Fortunately, it has been known for the model of Eqs.~\eqref{eq:H0p} and \eqref{eq:Hbare} that effective attraction is indeed found in anisotropic channels at the two-loop order~\cite{Twoloop}.
Here, we carry out the calculation explicitly (see Methods and Supplementary Note 2)
and find the effective interactions in the $|\tilde{l}|=1$ channel to be attractive, i.e., 

\begin{align}
\lambda_{\tau,\tau'}^{(0),|\tilde{l}|=1}=\frac{1}{\pi}\int_0^{\pi} d\theta g_{\tau,\tau'}^{(0)}(\theta)\Phi_{1}(\theta)<0
\label{eq:lambdadf}
\end{align}
for $\tau,\tau'=\uparrow,\downarrow$, where $\theta\equiv 2\sin^{-1}(\frac{|\textbf{q}\pm\textbf{q}'|}{2q_F})$ is the angle associated with the momentum transfer, and $\Phi_1(\theta)=\sqrt{2}\cos (\theta)$ is the normalized angular-momentum-one eigenstate in 2D. 

\begin{figure}[!hp]
\begin{centering}
\includegraphics[width=0.8\linewidth]{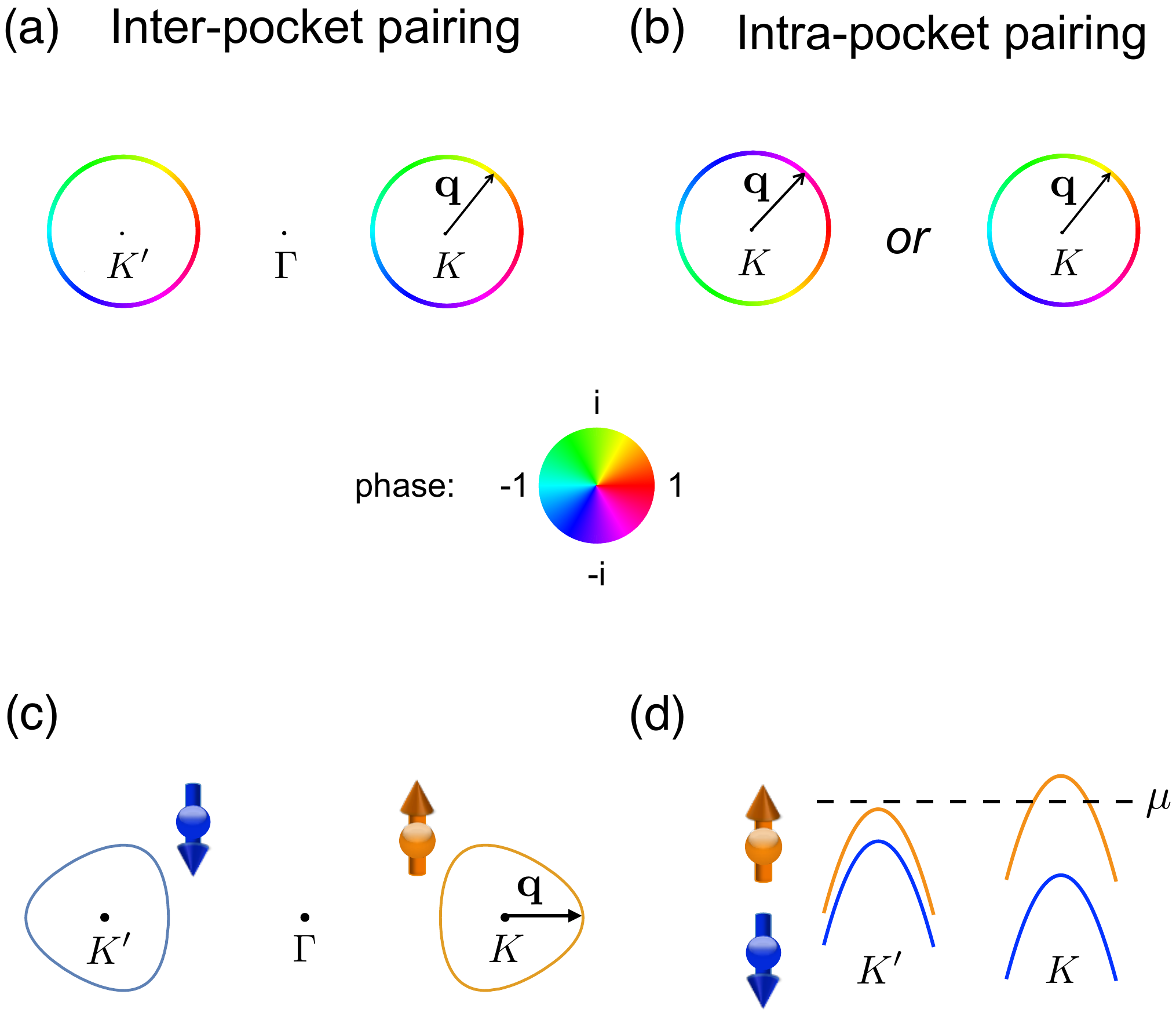}
\end{centering}
\caption{\textbf{The inter- and intra-pocket $|\tilde{l}|=1$ paired states.}
The gap functions of the $\tilde{l}=\pm 1$ paired states have the approximate form $q_x\pm iq_y$ on the two pockets (represented by hollow circles) centered at $\pm\textbf{K}$ which we assume to be small and circular as discussed in the text. The color scheme on the circles represents the phase of the gap functions, as indicated by the color wheel. 
(a) For the inter-pocket pairing case, the phase winding on the two pockets are locked to each other. Overall, the paired state breaks time-reversal symmetry. (b) For the intra-pocket pairing case, each pocket can independently have either  
$\tilde{l}=1$ or $\tilde{l}=-1$, which leads to a counterclockwise or clockwise phase winding of $2\pi$.
The possible factor and way to tilt the balance between the inter- and intra-pocket pairings: 
(c) A sketch for the trigonally warped Fermi pockets expected upon a heavier doping where the chemical potential still lies within the spin-split. Such trigonal warping is expected to suppress the intra-pocket pairing as an electron at $\textbf{q}$ has no pairing partner on the same pocket at $-\textbf{q}$. 
(d) The schematic low-energy dispersion near the two valleys for a monolayer TMD grown on a ferromagnetic substrate. As the chemical potential $\mu$ (represented by the dashed line) intersects only one band near one valley, the intra-pocket pairing is expected to be promoted.
}
\label{orderpara}
\end{figure}

In the low energy limit, the effective attractions in the $|\tilde{l}|=1$ channel at the intermediate energy scale $\Lambda_0$ in Eq.~\eqref{eq:lambdadf} will lead to the following two degenerate topological paired states (see Methods):
the inter-pocket $(p/d)$-wave pairing which is expected to be chiral [see Fig.~\ref{orderpara}(a)] and the modulated intra-pocket pairing [see Fig.~\ref{orderpara}(b)]. 
The degeneracy is expected for the model of Eqs.~\eqref{eq:H0p} and \eqref{eq:Hbare} with its rotational symmetry in the pseudo spin $\tau$. There are two ways this degeneracy can be lifted. Firstly, the trigonal warping will suppress intra-pocket pairing as the two points on the same pocket with opposing momenta will not be both on the Fermi surface any more [see Fig.~\ref{orderpara}(c)]. On the other hand, a ferromagnetic substrate which will introduce an imbalance between the two pockets which promotes intra-pocket pairing\cite{TMDFMbilayer} [see Fig.~\ref{orderpara}(d)]. 

\section{Discussion}  
\begin{figure}[!hp]
\begin{centering}
\includegraphics[width=0.9\linewidth]{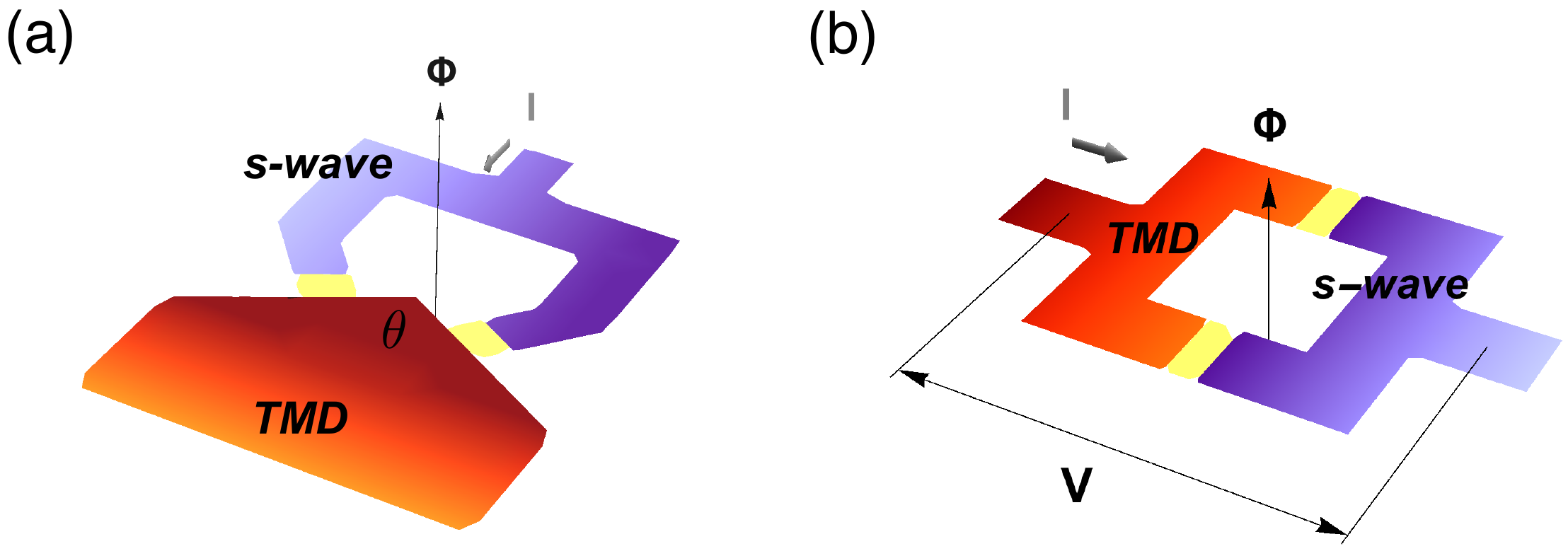}
\end{centering}
\caption{
\textbf{Configurations of possible SQUID experiments for probing the two paired states.}
In both (a) and (b), the red and blue parts indicate the lightly p-doped monolayer TMD and a uniform s-wave superconductor respectively, which are connected by two Josephson junctions represented by the yellow strips. $I$ is the applied current and $\Phi$ is the magnetic flux through the loop.
(a) shows the proposed dc SQUID interferometer set-up which can detect the anisotropy of the inter-pocket pairing symmetry. The flux-dependence of the critical current is expected to be insensitive and sensitive to the angle $\theta$ between the edges connected to the two junctions 
for isotropic and anisotropic pairing respectively. 
(b) shows the proposed dc SQUID interferometer set-up which can probe the finite pair-momentum of the intra-pocket pairs for the $C=0$ case. 
The TMD is oriented in the direction such that the phase of the pairing wavefunction is spatially modulated along the junction. 
The period in flux $\Phi$ of the modulated voltage $V$ across the SQUID loop is expected to be halved into $\frac{hc}{4e}$ since 
the difference between the pair-momenta on the two sides of a junction requires simultaneous tunneling of a spin-up and a spin-down intra-pocket pair, each carrying pair-momentum $2\textbf{K}$ and $-2\textbf{K}$, into the uniform superconductor. 
}
\label{detect}
\end{figure}

The distinct topological properties of the two predicted exotic superconducting states lead to unusual signatures. 
The inter-pocket $|\tilde{l}|=1$ paired state [see Fig.~\ref{orderpara}(a)] is topological with Chern number $|C|=2$ due to the two pockets (see Methods). The Chern number dictates for two chiral edge modes, which in this case are Majorana chiral edge modes each carrying central charge $\frac{1}{2}$\cite{CentralCharge,ReadGreen}. This is in contrast to $d+id$ paired state on a single spin-degenerate pocket which is another chiral superconducting state \cite{TwostepRGprb,Nandkishore2012,fRGdid,didTJSchaffer,SrPtAsFischer,didscNaCoO2} with four chiral Majorana edge modes. 
An unambiguous signature of two Majorana edge modes in the inter-pocket chiral $|\tilde{l}|=1$ paired state 
will be a quantized thermal Hall conductivity\cite{ReadGreen} of

\begin{align}
K_H=c\frac{\pi^2k_B^2}{3h}T
\label{eq:deltainter}
\end{align}
at temperature $T$, where $c=1$ is the total central charge. 
Additionally, signatures of the chiral nature of such state could be revealed by a detection of time-reversal symmetry breaking in polar Kerr effect and muon spin relaxation measurements.
Finally, a sharp signature of anisotropy of the pairing will be the maximization of the critical current in a direct current superconducting quantum interference device (dc SQUID) interferometry setup of Fig.~\ref{detect}(a) at some finite flux $\Phi_{max}\neq0$. 

The intra-pocket $|\tilde{l}|=1$ paired state [see Fig.~\ref{orderpara}(b)] is not only topological, but also its phase of the gap is spatially modulated with $e^{i2\textbf{K}\cdot\textbf{r}}$ and $e^{-i2\textbf{K}\cdot\textbf{r}}$ for spin-up and -down pairs respectively, where $\textbf{r}$ is the spatial coordinate of the center of mass of the pair (see Supplementary Note 3).  
Since the gaps on the two pockets are not tied to each other in principle,
the system may be either helical respecting time-reversal symmetry ($C=0$) or chiral ($C=2$). Either way, there will be a Majorana zero mode of each spin species at a vortex core so long as $\tau_z$ is preserved.
What makes the intra-pocket paired state distinct from existing candidate materials for topological superconductivity, however, is its spatial modulation.
Smoking gun signature of the modulation in phase would be the halved period $\frac{hc}{4e}$ of the oscillating voltage across the dc SQUID setup in Fig.~\ref{detect}(b) in flux $\Phi$ due to the difference between the pair-momenta on the two sides of the junction.  
Another signature of the intra-pocket paired state will be the spatial profile of the modulated phase directly detected with an atomic resolution scanning Josephson tunneling microscopy (SJTM)\cite{ScanJTDynes08,SJTMpdwDavis}. 

In summary, we propose the ${\textbf k}$-space spin splitting as a new strategy for 
topological superconductivity.
Specifically, we predict lightly p-doped monolayer TMDs with their spin-valley-locked band structure and correlations to exhibit topological superconductivity.
Of the monolayer TMDs, $\rm{WSe_2}$ may be the most promising as its large spin-splitting energy scale\cite{ARPESWSe2} allows for substantial carrier density within the spin-valley-locked range of doping\cite{EffMassTMD}. 
The rationale for the proposed route is to use a lower symmetry to restrict the pairing channel. The merit of this approach is clear when we contrast the proposed setting to the situation of typical spin-degenerate trigonal systems. With a higher symmetry, trigonal systems typically deals with the need for anisotropic pairing due to the repulsive interaction by turning to the topologically trivial $f$-wave channel 
\cite{TwostepRGprb,ChiralSCreviewKallin}. 
The n-doped TMDs 
whose low-energy band structure is approximately spin-degenerate fall into this category. Hence, experimentally realized superconductivity in n-doped systems would likely be topologically trivial even if the superconductivity is driven by the same repulsive interaction we consider here.
The predicted topological paired states in p-doped TMDs are a direct consequence of the spin-valley locking which breaks the spin-degeneracy in {$\textbf k$} space and creates two species of spinless fermions. Experimental confirmation of the predicted topological superconductivity in p-doped TMDs will open unprecedented opportunities in these highly tunable systems. 

\section{Methods}

\textbf{Perturbative renormalization group (RG) calculation}\\
For the RG calculation, we follow the perturbative two-step RG procedure in Ref. ~\onlinecite{TwostepRGprb}, which has been used to study superconductivity in systems such as $\rm{Sr_2RuO_4}$\cite{RaghuPRL} and generic hexagonal lattices with spin-degeneracy\cite{TwostepRGprb}. 
Taking the Hubbard on-site repulsion in Eq. (2) as the microscopic interaction, the first step is to integrate out higher energy modes and obtain $g^{(0)}_{\tau,\tau'}$ in Eq. (3), the low-energy effective interactions in the Cooper channel at an intermediate energy $\Lambda_0\gtrsim 0$ close to the Fermi level. The second step is to study the evolution of these effective interactions as the energy flows from $\Lambda_0$ to $0$, which is governed by the RG equations.  

In the first step, we calculate the inter- and intra-pocket effective interactions $g_{\rm{inter}}^{(0)}(\textbf{q},\textbf{q}')\equiv g^{(0)}_{\tau,\bar{\tau}}(\textbf{q},\textbf{q}')$ and $g_{\rm{intra}}^{(0)}(\textbf{q},\textbf{q}')\equiv g^{(0)}_{\tau,\tau}(\textbf{q},\textbf{q}')$ in terms of the incoming and outgoing momenta $\textbf{q}$ and $\textbf{q}'$ order by order in $U$ until we obtain attraction in one of them in certain partial-wave channel $\tilde{l}$. 
Following Ref.~ \onlinecite{Twoloop}, we find the effective interactions to be (see Supplementary Note 2)

\begin{align}
g_{\rm{inter}}^{(0)}(\textbf{q},\textbf{q}')
\sim C+\frac{m^2U^3}{2\pi^3}\frac{\sqrt{4q_F^2-p'^2}}{2q_F}-\frac{U^3m^2}{64\pi^3}(1-\frac{p^2}{4q_F^2}) \log [1-\frac{p^2}{4q_F^2}],
\label{eq:ginterall}
\end{align}
and 

\begin{align}
g_{\rm{intra}}^{(0)}(\textbf{q},\textbf{q}')
\sim C'-\frac{m^2U^3}{2\pi^3}\frac{\sqrt{4q_F^2-p^2}}{2q_F}-\frac{U^3m^2}{64\pi^3}(1-\frac{p^2}{4q_F^2}) \log [1-\frac{p^2}{4q_F^2}],
\label{eq:gintraall}
\end{align}
where $\textbf{p}=\textbf{q}\pm\textbf{q}'$ is the momentum transfer, $C>0$ and $C'<0$ are momentum-independent constants coming from tree level and one-loop order, and the momentum-dependent terms come solely from two-loop order.

Each partial-wave $\tilde{l}$ component is given by the projection of $g_{\rm{inter}/\rm{intra}}^{(0)}(\textbf{q},\textbf{q}')$ on to the normalized angular momentum $\tilde{l}$ eigenstate in 2D, $\Phi_{\tilde{l}}(\theta)=\sqrt{2}\cos \tilde{l}\theta$, where $\theta\equiv 2\sin^{-1}(\frac{p}{2q_F})$ is the angle associated with the momentum transfer $\textbf{p}$. We find 

\begin{align}
\lambda_{\rm{inter}/\rm{intra}}^{(0),\tilde{l}}&=\frac{1}{\pi}\int_0^{\pi} d\theta g_{\rm{inter}/\rm{intra}}^{(0)}(\theta)\Phi_{\tilde{l}}(\theta)
=\frac{2\sqrt{2}\alpha}{\pi}\frac{(\pm1)^{\tilde{l}+1}}{1-4\tilde{l}^2}
-\frac{\beta}{\sqrt{2}\pi} \frac{H_{1-\tilde{l}}+H_{1+\tilde{l}}+2\log2-3}{\tilde{l}(1-\tilde{l}^2)}\sin(\tilde{l}\pi),
\label{eq:lambdafull}
\end{align}
where $H_n$ is the $\rm{n^{th}}$ harmonic number, and $\alpha\equiv\frac{U^3m^2}{2\pi^3}$ and $\beta\equiv \frac{U^3m^2}{64\pi^3}$ are postive constants related to density of states and interaction strength. 
Here, terms with $\alpha$ and $\beta$ come from contributions with one particle-particle and one particle-hole bubble, and two particle-hole bubbles, respectively (see Supplementary Note 2). 
The $\alpha$ term in $\lambda_{\rm{intra}}^{(0),\tilde{l}}$ acquires an extra minus sign on top of $(-1)^{\tilde{l}}$ from the closed fermion loops in Supplementary Fig.1 (3g) and (3h). 
Meanwhile, the $\alpha$ term in $\lambda_{\rm{inter}}^{(0),\tilde{l}}$ contains an implicit $(-1)^{\tilde{l}}$ factor due to the fact that the outgoing external momenta in Supplementary Fig.1 (3a) and (3b) are exchanged, which is equivalent to setting $\Phi_{\tilde{l}}(\theta)\rightarrow \Phi_{\tilde{l}}(\pi-\theta)$.

Note that $\lambda_{\rm{intra}}^{(0),\tilde{l}}$ with even $\tilde{l}$s are forbidden since intra-pocket pairs have equal spin, and that $\lambda_{\rm{inter}}^{(0),\tilde{l}}=\lambda_{\rm{intra}}^{(0),\tilde{l}}$ for odd $\tilde{l}$s since they correspond to the spin-triplet states with $\tau_z=0$ and $\pm 1$ respectively.
While $\lambda^{(0),0}_{\rm{inter}}>0$ as expected from the bare repulsion, the most negative values are  $\lambda_{\rm{inter}}^{(0),\pm1}=\lambda_{\rm{inter}}^{(0),\pm 1}\sim-0.3\alpha-0.04\beta<0$.  

In the second step, we derive and solve the RG equations to study the evolutions of the effective interactions $\lambda_{\rm{inter/intra}}^{\tilde{l}}(E)$ as the energy $E$ lowers from $\Lambda_0$ to $0$. Using $\lambda^{(0),\tilde{l}}_{\rm{inter}/\rm{intra}}$ in Eq.~\eqref{eq:lambdafull} as the initial values for the RG flows, the channel with the most relevant attraction in the low-energy limit $E\rightarrow 0$ is the dominant pairing channel. 
Under the assumption that the energy contours for $0<E<\Lambda_0$ are isotropic, different partial-wave components do not mix while the inter- and intra-pocket interactions with the same $\tilde{l}$ can in principle mix. 
By a procedure similar to that in Ref.~ \onlinecite{Nandkishore2012} and ~\onlinecite{ChubukovRGreview}, we find the RG equations up to one-loop order to be

\begin{align}
\frac{d\lambda_{\rm{inter}}^{\tilde{l}}}{dy}=-(1-d_2)(\lambda_{\rm{inter}}^{\tilde{l}})^2
\label{eq:RGeqninter}
\end{align} 
and 

\begin{align}
\frac{d\lambda_{\rm{intra}}^{\tilde{l}}}{dy}=-(d_1-d_3)(\lambda_{\rm{intra}}^{\tilde{l}})^2-2d_3(\lambda_{\rm{inter}}^{\tilde{l}})^2, 
\label{eq:RGeqnintra}
\end{align} 
where the inverse energy scale $y\equiv\Pi^{s\bar{s}}_{pp}(0)\sim\nu_0\log(\Lambda_0/E)$ is the RG running parameter,  $d_1(y)\equiv\frac{\partial\Pi_{pp}^{ss}(\pm2\textbf{K})}{\partial y}$, $d_2(y)\equiv\frac{\partial\Pi_{ph}^{s\bar{s}}(\pm2\textbf{K})}{\partial y}$, and $d_3(y)\equiv\frac{\partial\Pi_{ph}^{ss}(0)}{\partial y}$. Here, $\Pi_{pp/ph}^{ss'}(\textbf{k})$ is the non-interacting static susceptibility at momentum $\textbf{k}$ in the particle-particle or particle-hole channel defined in  Supplementary Note 1. Since the low-energy band structure is well-nested at $\pm2\textbf{K}$ in the particle-particle channel, the Cooper logarithmic divergence appears not only at $\textbf{k}=0$ but also $\pm2\textbf{K}$ (see Supplementary Note 1). Thus, 
$d_1(y)=1$. On the other hand, since the low-energy band structure is poorly-nested at any $\textbf{k}$ in the particle-hole channel and is far from van Hove singularity, 
the particle-hole susceptibilities do not diverge in the low-energy limit (see Supplementary Note 1). Thus, $d_2(y), d_3(y)\ll 1$ in the low-energy limit $y\rightarrow\infty$. 
Therefore with logarithmic accuracy, the inter- and intra-pocket interactions renormalize independently with the well-known RG equation in the Cooper channel

\begin{align}
\frac{d\lambda^{\tilde{l}}_{i}}{dy}=-(\lambda_i^{\tilde{l}})^2
\label{eq:RGeqn}
\end{align} 
with $i=$ inter, intra. 
The RG flow
$\lambda_i^{\tilde{l}}(y)=\frac{\lambda_i^{(0),\tilde{l}}}{1+\lambda_i^{(0),\tilde{l}}y}$ which solves the RG equation shows that the pairing interaction in channel $\tilde{l}$ becomes a marginally relevant attraction only if the initial value $\lambda_i^{(0),\tilde{l}}<0$. 
Since we concluded that the most negative initial values occur in the $|\tilde{l}|=1$ channels for both inter- and intra-pocket interactions in the first step of the RG procedure, we expect degenerate inter- and intra-pocket $|\tilde{l}|=1$ pairings in the low-energy limit.\\

\textbf{The Chern number of inter-pocket paired state}\\
The inter-pocket chiral $|\tilde{l}|=1$ paired state becomes just a spinful $p+ip$ paired state with total spin $\tau_z=0$ when we map the spin-valley-locked two-pocket problem to a spin-degenerate singlet-pocket problem. 
The spinful $p+ip$ pairing comprises two copies of `spinless' $p+ip$ pairings as the Bogoliubov-de Gennes (BdG) Hamiltonian of the former can be written as

\begin{align}
H&=\sum_{\textbf{q}} \epsilon_{\textbf{q}}(c^{\dagger}_{\textbf{q},\uparrow}c_{\textbf{q},\uparrow}+c^{\dagger}_{\textbf{q},\downarrow}c_{\textbf{q},\downarrow})+
\Delta_{\textbf{q}}(c^{\dagger}_{\textbf{q},\uparrow}c^{\dagger}_{-\textbf{q},\downarrow}+c^{\dagger}_{\textbf{q},\downarrow}c^{\dagger}_{-\textbf{q},\uparrow})+H.c.\nonumber\\
&=\sum_{\textbf{q}} (\epsilon_{\textbf{q}}c^{\dagger}_{\textbf{q},+}c_{\textbf{q},+}
+\Delta_{\textbf{q}}c^{\dagger}_{\textbf{q},+}c^{\dagger}_{-\textbf{q},+}+H.c.)+(\epsilon_{\textbf{q}}c^{\dagger}_{\textbf{q},-}c_{\textbf{q},-}
-\Delta_{\textbf{q}}c^{\dagger}_{\textbf{q},-}c^{\dagger}_{-\textbf{q},-}+H.c.),
\label{eq:interBdG}
\end{align} 
where the low-energy dispersion $\epsilon_{\textbf{q}}=-\frac{q^2}{2m}-\mu$, the gap function $\Delta_{\textbf{q}}\sim q_x\pm iq_y$, and $c_{\textbf{q},\pm}\equiv(c_{\textbf{q},\uparrow}\pm c_{\textbf{q},\downarrow})/\sqrt{2}$. 
Since a spinless $p+ip$ paired state has Chern number $C=1$, where $C=\frac{1}{8\pi}\int d^2q~\hat{\textbf{m}}\cdot [\partial_{q_x}\hat{\textbf{m}}\times\partial_{q_y}\hat{\textbf{m}}]$ with $\hat{\textbf{m}}=(\rm{Re}[\Delta_{\textbf{q}}],~\rm{Im}[\Delta_{\textbf{q}}],~\epsilon_{\textbf{q}})/\sqrt{\epsilon_{\textbf{q}}^2+|\Delta_{\textbf{q}}|^2}$, 
 the $\tau_z=0$ spinful $p+ip$ paired state in the single-pocket system has $C=2$. Hence, the inter-pocket chiral $|\tilde{l}|=1$ pairing in the two-pocket system has $C=2$ as well. \\

\textbf{Data Availability Statement}
The authors declare that the data supporting the findings of this study are available within the paper and its Supplementary Information file.


\newpage
\textbf{Acknowledgements}\\
The authors thank Reza Asgari, Debdeep Jena, Katja Nowak, Grace Xing, K. T. Law, and Andrey Chubukov for helpful discussions. Y.-T.H. and E.-A.K. were supported by the Cornell Center for Materials Research with funding from the NSF MRSEC program (DMR-1120296). E.-A.K. was supported in part by by the National Science Foundation (Platform for the Accelerated Realization, Analysis, and Discovery of Interface Materials (PARADIM)) under Cooperative Agreement No. DMR-1539918. A. V. was supported by Gordon and Betty Moore Foundation and in part by Bethe postdoctoral fellowship. MHF acknowledges support from the Swiss Society of Friends of the Weizmann Institute.\\

\textbf{Author Contributions}\\
Y.-T.H. carried out the RG calculations to identify the dominant paired states. 
Y.-T.H. and A.V. analyzed the topological properties for the paired states.
Y.-T.H. and MHF analyzed the pairing symmetries for the paired states.
E.-A.K supervised the project and wrote the paper with contributions from Y.-T.H., A.V., and MHF.

* To whom correspondence should be addressed:  eun-ah.kim@cornell.edu\\

\textbf{Conflict of interest statement}\\
The authors declare no competing financial interests.

\newpage

\newpage

\newpage
{\bf\normalsize{Supplementary Note 1: Why only pairing instabilities?}}\\
In this work we consider only pairing instabilities but not particle-hole instabilities, e.g. spin density waves. 
In the following, we explain why the pairing instabilities are expected to dominate over particle-hole instabilities in the low energy limit. 
Whether instabilities in particle-hole channel or particle-particle (pairing) channel dominate depends on which of the non-interacting static susceptibilities in particle-hole channel $\Pi_{ph}(\textbf{p})$ and particle-particle channel $\Pi_{pp}(\textbf{p})$ diverges faster as approaching the low energy limit. 
These susceptibilities of electrons with spin $s$ and low-energy dispersion $\epsilon^{s}(\textbf{k})$ have the form
\begin{align} 
\Pi_{pp}^{ss'}(\textbf{p})\equiv
\sum_{n}\int \frac{d^2k}{4\pi^2}G^{s}(i\omega_{n},\textbf{k})G^{s'}(-i\omega_{n},-\textbf{k}+\textbf{p})
=\int\frac{d^2k}{(2\pi)^2}\frac{1-f(\epsilon^s_{-\textbf{k}+\textbf{p}})-f(\epsilon^{s'}_{\textbf{k}})}{\epsilon^s(-\textbf{k}+\textbf{p})+\epsilon^{s'}(\textbf{k})}
\label{eq:Pippdf}
\end{align} 
and
\begin{align} 
\Pi_{ph}^{ss'}(\textbf{p})\equiv
-\sum_{n}\int \frac{d^2k}{4\pi^2}G^{s}(i\omega_{n},\textbf{k})G^{s'}(i\omega_{n},\textbf{k}+\textbf{p})
=-\int\frac{d^2k}{(2\pi)^2}\frac{f(\epsilon^s_{\textbf{k}+\textbf{p}})-f(\epsilon^{s'}_{\textbf{k}})}{\epsilon^s(\textbf{k}+\textbf{p})-\epsilon^{s'}(\textbf{k})},
\label{eq:Piphdf}
\end{align} 
where spin $s,s'=\uparrow/\downarrow$, $\omega_{n}$ is the fermionic Matsubara frequency, $\textbf{k}$ and $\textbf{p}$ are momenta, $G^{s}(i\omega_{n},\textbf{k})=\frac{1}{i\omega_n-\epsilon^s(\textbf{k})}$ is the non-interacting Green's function, and $f(\epsilon^s_{\textbf{k}})$ is the Fermi function at temerature $T$. 

In general, $\Pi_{pp}^{ss'}(\textbf{p})$ always diverges logarithmically at total-momentum $\textbf{p}=0$ despite the low-energy dispersion $\epsilon^s_{\textbf{k}}$, which indicates pair-momentum 0 superconductivity if dominates. 
On the other hand, $\Pi_{ph}^{ss'}(\textbf{p})$ typically diverges at momentum-transfer $\textbf{p}=0$ when the density of states diverges, i.e. near the van Hover singularity, or at some finite momentum-transfer $\textbf{p}=\textbf{Q}$ when the Fermi surface is nested in the particle-hole channel at $\textbf{Q}$. The former and latter indicate instabilities such as ferromagnetism and density-waves respectively when they each dominates. In a two-pocket system, this requires a hole and an electron pocket to have the same low-energy dispersion (but opposite in energy). 
In the case where susceptibilities in the two channels diverge equally fast, one needs to further compare their corresponding driving interactions to determine the dominant instability.

In the current lightly p-doped monolayer TMD case, note that both pockets are hole pockets though they have the same low-energy dispersion $\epsilon^{\uparrow}(\textbf{k})=-\frac{(\textbf{k}-\textbf{K})^2}{2m}$ and $\epsilon^{\downarrow}(\textbf{k})=-\frac{(\textbf{k}+\textbf{K})^2}{2m}$ with respect to their own valley centers $\textbf{K}$ and $-\textbf{K}$ upon low-doping. Thus, the Fermi surface is in fact poorly nested at $2\textbf{K}$ in the particle-hole channel. To be precise, since $\epsilon^{\downarrow}(\textbf{p})=\epsilon^{\uparrow}(\textbf{p}+2\textbf{K})$, the particle-hole susceptibility has the relation
\begin{align} 
\Pi_{ph}^{s\bar{s}}(2\textbf{K}+\textbf{p})=-\int \frac{d^2k}{4\pi^2}\frac{f^{s}_{\textbf{k}+2\textbf{K}+\textbf{p}}-f^{\bar{s}}_{\textbf{p}}}{\epsilon^{s}(\textbf{k}+2\textbf{K}+\textbf{p})-\epsilon^{\bar{s}}(\textbf{k})}
=-\int \frac{d^2k}{4\pi^2}\frac{f^{\bar{s}}_{\textbf{k}+\textbf{p}}-f^{\bar{s}}_{\textbf{k}}}{\epsilon^{\bar{s}}(\textbf{k}+\textbf{p})-\epsilon^{\bar{s}}(\textbf{k})}=\Pi_{ph}^{\bar{s}\bar{s}}(\textbf{p})
\label{eq:Piph}
\end{align} 
for $s=\uparrow$ and $\bar{s}=-s$. 
Thus, 
\begin{align}
\Pi_{ph}^{s\bar{s}}(\pm2\textbf{K})=\Pi_{ph}^{ss}(0)\sim\nu_0 
\label{eq:Piph0}
\end{align}
is not diverging in the low energy limit as long as the density of states on the Fermi surface $\nu_0$ is finite. 
Therefore, we do not consider particle-hole susceptibilities in this work. 

On the other hand, since the Fermi surface is perfectly nested at $2\textbf{K}$ in the particle-particle channel, the particle-particle susceptibility 
\begin{align}
\Pi_{pp}^{s\bar{s}}(0)=\Pi_{pp}^{ss}(\pm2\textbf{K})
\sim\nu_0{\rm{Log}}(\frac{\Lambda}{E}) 
\label{eq:Pipp0}
\end{align}
diverges logarithmically as approaching the low-energy limit  $E\rightarrow 0$ with $\Lambda$ being the UV cutoff scale. Note that the Cooper logarithmic divergence does not occur only at the usual $\textbf{p}=0$, but also at $\textbf{p}=2\textbf{K}$. This indicates that the superconductivity with pair momentum $0$ (spatially uniform) and $2\textbf{K}$ (spatially modulated at 2$\textbf{K}$) could be equally dominant in the low energy. To determine which is truly more dominant, we need to study their pairing interactions using the RG analysis in the following section.\\

{\bf\normalsize{Supplementary Note 2: Inter- and intra-pocket effecitve interactions}}\\

\begin{figure}[]
\centering
\includegraphics[width=14cm]{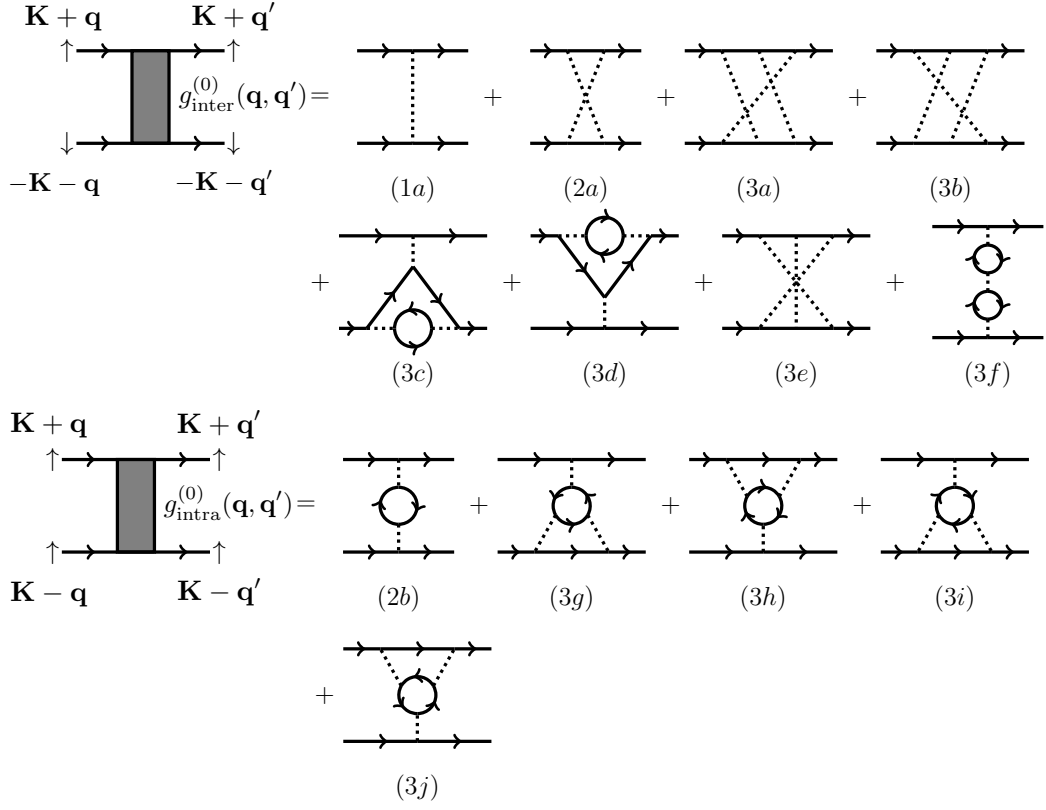}
\caption{Feymann diagrams for the contributions up to two-loop order to the inter- and intra-pocket effective interactions $g_{\rm{inter/intra}}^{(0)}(\textbf{q},\textbf{q}')$ at the intermediate energy scale $E=\Lambda_0$. The solid and dotted lines represent fermions and repulsive Hubbard interaction $U$ respectively.}
\label{diagram}
\label{diagram}
\end{figure}


We will calculate the inter- and intra-valley effective interactions $g_{\rm{inter}}^{(0)}(\textbf{q},\textbf{q}')\equiv g^{(0)}_{\tau,\bar{\tau}}(\textbf{q},\textbf{q}')$ and $g_{\rm{intra}}^{(0)}(\textbf{q},\textbf{q}')\equiv g^{(0)}_{\tau,\tau}(\textbf{q},\textbf{q}')$ at energy $\Lambda_0$ in terms of the incoming and outgoing momenta $\textbf{q}$ and $\textbf{q}'$ order by order in $U$ until we obtain attraction in one of them in certain partial-wave channel $\tilde{l}$. 
Before we start, notice that by omitting the valley index, which is inter-locked with the spin index $\tau$ for the low-energy fermions, the inter- and intra-valley interactions in the spin-valley locked two-pocket picture are just the opposite- and equal-spin interactions in a spin-degenerate single-pocket picture. 
Fortunately, Ref. ~\onlinecite{Twoloop} has already studied the pairing problem in a spin-degenerate single-pocket system under repulsive Hubbard interaction described by Eq. (1) and Eq. (2). 
Thus, we expect the same result as Ref. ~\onlinecite{Twoloop}, i.e. the largest attraction occuring in the angular-momentum-one channel, but with a different physical meaning when mapping back to the spin-valley locked two-pocket picture. 
To make the mapping between the two pictures explicit, we will follow Ref. ~\onlinecite{Twoloop} to calculate
$g_{\rm{inter}}^{(0)}(\textbf{q},\textbf{q}')$ and $g_{\rm{intra}}^{(0)}(\textbf{q},\textbf{q}')$
in the two-pocket picture and denote the spin $s$ and valley $\pm\textbf{K}$ separately.\\ 

{\textbf{2.1 Tree level}}\\

At the tree level, the on-site repulsion $U>0$ only contribute to the inter-pocket interaction because $U$ acts between only electrons with opposite spins due to Pauli exclusive principle. Thus, 
\begin{align}
g_{\rm{inter}}^{(0),1}(\textbf{q},\textbf{q}')= U 
\label{eq:ginter1}
\end{align}
with the superscript $1$ denoting the tree-level contribution [see Supplementary Fig. \ref{diagram}(1a)]. 
This bare repulsion contributes to only the $\tilde{l}=0$ component of $g_{\rm{inter}}^{(0)}$ because $U$ is independent of the incoming and outgoing momenta $\textbf{q}$ and $\textbf{q}'$. 
Since this is a perturbative analysis, the inter-pocket $\tilde{l}=0$ pairing is suppressed regardless what the higher order contributions to $\tilde{l}=0$ channel are. To have any finite contribution to aisotropic channels ($\tilde{l}\neq 0$) requires momentum-dependence from loop corrections.\\

{\textbf{2.2 Second order}}\\
 
The $U^2$ (one-loop) order contributions to inter- and intra-pocket interactions are 
\begin{align}
g_{\rm{inter}}^{(0),2}(\textbf{q},\textbf{q}')= U^2\Pi_{ph}^{s\bar{s}}(\pm\textbf{Q}+\textbf{q}+\textbf{q}'),
\label{eq:ginter2}
\end{align}
and 
\begin{align}
g_{\rm{intra}}^{(0),2}(\textbf{q},\textbf{q}')=-U^2\Pi_{ph}^{ss}(\textbf{q}-\textbf{q}')
\label{eq:gintra2}
\end{align}
respectively for $s=\uparrow/\downarrow$, where the superscript $2$ denotes corrections from the second order [see Supplementary Fig. \ref{diagram}(2a) and (2b)].  
The particle-hole susceptibilies defined in Supplementary Eq. \eqref{eq:Piphdf} can be calculated as 
\begin{align}
\Pi_{ph}^{ss}(\textbf{p})=\frac{m}{2\pi^2}\int_{-\pi/2}^{\pi/2} d\phi \int_{k_2}^{k_1} dk \frac{k}{kp\cos\phi}=\frac{m}{2\pi}=\Pi_{ph}^{\bar{s}s}(\textbf{p})
\label{eq:Piphconst1}
\end{align}
where $k_{1/2}\equiv \pm\frac{p}{2}\cos\phi+\frac{\sqrt{4q_F^2-p^2\sin^2\phi}}{2}$. 
Thus, the one-loop corrections are still momentum-independent and contribute to only the $\tilde{l}=0$ channel. This is a consequence of isotropic parabolic dispersion in 2D\cite{Twoloop,TwostepRGprb}.  
Note that though $g_{\rm{intra}}^{(0),2}$ seems to imply $\tilde{l}=0$ intra-pocket pairing, pairings in even $\tilde{l}$ channels are not allowed since these are equal-spin pairs.
Thus, if either the inter- or intra-pocket pairing were to occur at all, the effective attraction has to come from at least two-loop order.\\

{\textbf{2.3 Third order}}\\

The $U^3$ (two-loop) contribution of short-range repulsion for a spin-degenerate rotational-invariant 2D system with a single pocket and parabolic dispersion has been proven to facilitate p-wave (angular momentum 1) pairing\cite{Twoloop}. Since both pockets in p-doped TMDs have the same low-energy effective dispersion which is parabolic upon light doping, we can map this spin-valley locked two-pocket system to the spin-degenerate single-pocket system studied in Ref. \onlinecite{Twoloop} by bringing the pocket centers K and K' both to $\textbf{k}=0$. 
Thus, we expect to obtain the largest attractions in the partial-wave channel $\tilde{l}=1$ as well, but the partial-wave channels here are with respect to K and K' instead of $\Gamma$. This indicates degenerate inter- and intra-pocket pairings with $\tilde{l}=1$ after we map back to the two-pocket system. 
In the following, we will show the calculations of $g_{\rm{intra}/\rm{inter}}^{(0)}$ following Ref.~ \onlinecite{Twoloop} to comfirm our expectation. 

From the corresponding diagrammatic expressions shown in Supplementary Fig. \ref{diagram}(3a)-(3j), we can see that the two-loop contributions can be divided into two groups: the ones with one particle-particle and one particle-hole bubble (diagram 3a, 3b, 3g and 3h), and the ones with two particle-hole bubbles (diagram 3c, 3d, 3e, 3f, 3i, and 3j).
We first calculate the former contributions to intra-pocket interaction, i.e. diagram (3g) and (3h).
In the static limit, 
\begin{align}
g_{\rm{intra}}^{pp}(\textbf{q},\textbf{q}')&=g_{3g}(\textbf{q},\textbf{q}')+g_{3h}(\textbf{q},\textbf{q}')\nonumber\\
&=-U^3\sum_{n,\tilde{n}}\int \frac{d^2l}{4\pi^2}\int \frac{d^2\tilde{l}}{4\pi^2}G^{\uparrow}(i\omega_{\tilde{n}},\tilde{\textbf{l}})G^{\downarrow}(-i\omega_{\tilde{n}},-\tilde{\textbf{l}}+\textbf{t}_+)G^{\downarrow}(i\omega_{n},\textbf{l}+\frac{\textbf{p}}{2})G^{\downarrow}(i\omega_{n},\textbf{l}-\frac{\textbf{p}}{2})\nonumber\\
&+(\textbf{t}_+\rightarrow\textbf{t}_-)
\label{eq:gintrapp}
\end{align}
where $\omega_{n(\tilde{n})}$ is the fermionic Matsubara frequency, $G^s(i\omega_{n},\textbf{l})=\frac{1}{i\omega_{n}-\epsilon^{s}(\textbf{l})}$ is the non-interacting Green's function, $\textbf{t}_{\pm}\equiv \textbf{l}\pm\frac{\textbf{q}+\textbf{q}'}{2}$, $\textbf{p}\equiv\textbf{q}'-\textbf{q}$, and $\textbf{q}$ and $\textbf{q}'$ are the external incoming and outgoing momenta relative to the valley center $\textbf{K}$. As the electrons have energy $E=\Lambda_0\gtrsim 0$, $|\textbf{q}|=|\textbf{q}'|\sim q_F$ with $q_F$ being the Fermi momentum of a single pocket. The over-all minus sign results from the closed fermion loop. The particle-particle loop integral can be calculated as
\begin{align}
&\sum_{\tilde{n}}\int \frac{d^2\tilde{l}}{4\pi^2}G^{\uparrow}(i\omega_{\tilde{n}},\tilde{\textbf{l}})G^{\downarrow}(-i\omega_{\tilde{n}},-\tilde{\textbf{l}}+\textbf{t}_+)
=\int \frac{d^2\tilde{l}}{4\pi^2}\frac{1-f(\epsilon^{\uparrow}_{\tilde{\textbf{l}}})-f(\epsilon^{\downarrow}_{-\tilde{\textbf{l}}+\textbf{t}_+})}{\epsilon^{\uparrow}_{\tilde{\textbf{l}}}+\epsilon^{\downarrow}_{-\tilde{\textbf{l}}+\textbf{t}_+}}
=\int \frac{d^2\tilde{l}}{4\pi^2}\frac{1-f(\epsilon^{\uparrow}_{\tilde{\textbf{l}}})-f(\epsilon^{\uparrow}_{\tilde{\textbf{l}}-\textbf{t}_+})}{\epsilon^{\uparrow}_{\tilde{\textbf{l}}}+\epsilon^{\uparrow}_{\tilde{\textbf{l}}-\textbf{t}_+}}\nonumber\\
&=-2m\int \frac{d^2\tilde{l}}{4\pi^2}\frac{1-f(\epsilon^{\uparrow}_{\tilde{\textbf{l}}+\textbf{K}+\frac{\textbf{t}_+}{2}})-f(\epsilon^{\uparrow}_{\tilde{\textbf{l}}+\textbf{K}-\frac{\textbf{t}_+}{2}})}{(\tilde{\textbf{l}}+\frac{\textbf{t}_+}{2})^2+(\tilde{\textbf{l}}-\frac{\textbf{t}_+}{2})^2-2q_F^2}
=-\frac{2m}{4\pi^2}\int_{-\pi/2}^{\pi/2} d\tilde{\phi} (\int_{0}^{\tilde{l}_2}-\int_{\tilde{l}_1}^{r_0^{-1}})d\tilde{l}\frac{\tilde{l}}{\tilde{l}^2+\frac{t_+^2}{4}-q_F^2}\nonumber\\
&\sim -\frac{m}{2\pi}(\ln[\frac{t_+^2}{q_F^2}]+c_0)
\label{eq:gintrapp1}
\end{align}
assuming $t_{\pm}\ll2q_F$, which is the regime where the main momentum-dependence comes from\cite{Twoloop}. 
Here, 
$\tilde{\phi}$ is the angle between the loop momentum $\tilde{\textbf{l}}$ and $\textbf{t}_+$, $r_0^{-1}$ the UV cutoff for momentum integral, $\tilde{l}_{1/2}\equiv \pm\frac{t_+\cos\tilde{\phi}}{2}+ \frac{t_+}{2}\sqrt{\frac{4q_F^2}{t_+^2}-\sin^2\tilde{\phi}}$, and $c_0$ contains terms independent of $\textbf{t}_{\pm}$. We will drop $c_0$ in the following since our purpose is to obtain the momentum-dependent part. Plugging Supplementary Eq. \eqref{eq:gintrapp1} back to Supplementary Eq. \eqref{eq:gintrapp}, we obtain the second loop integral 
\begin{align}
g_{\rm{intra}}^{pp}(\textbf{q},\textbf{q}')&=\frac{+mU^3}{2\pi}\sum_{n}\int \frac{d^2l}{4\pi^2}\ln[\frac{t_+^2t_-^2}{q_F^4}]G^{\downarrow}(i\omega_{n},\textbf{l}+\frac{\textbf{p}}{2})G^{\downarrow}(i\omega_{n},\textbf{l}-\frac{\textbf{p}}{2})\nonumber\\
&=\frac{-2m^2U^3}{2\pi}\int\frac{d^2l}{4\pi^2}\ln[\frac{t_+^2t_-^2}{q_F^4}]\frac{f_{\textbf{l}+\frac{\textbf{p}}{2}}-f_{\textbf{l}-\frac{\textbf{p}}{2}}}{(\textbf{l}+\frac{\textbf{p}}{2})^2-(\textbf{l}-\frac{\textbf{p}}{2})^2}\nonumber\\
&\sim \frac{-m^2U^3}{4\pi^3}\int_{0}^{\pi/2} \frac{d\phi}{\cos\phi} \int_{\bar{l_2}}^{\bar{l_1}} d\bar{l}\ln[(\bar{l}^2-\epsilon^2)^2+4\epsilon^2\bar{l}^2\cos^2\phi]
\label{eq:gintrapp2}
\end{align}
where $\phi$ is the angle between $\textbf{p}$ and $\textbf{l}$, $\epsilon^2\equiv \frac{4q_F^2-p^2}{p^2}$, $\bar{l}\equiv \frac{2l}{p}$, and $\bar{l}_{1/2}\equiv \pm\cos\phi+\sqrt{\epsilon^2+\cos^2\phi}$. 
Here, $\epsilon \ll 1$ is a small parameter as we assumed $t_+/q_F\ll1$ in the first loop, which corresponds to the regime where the external momenta statisfy $p\sim 2q_F$ and the loop momentum $l/q_F\sim\bar{l}\ll1$. 
Notice that the integral is dominated by the regime of $\phi$ where $\cos\phi=O(\epsilon)$ is another small parameter besides $\epsilon$.  
Since we are interested in the portion of scattering amplitude which depends on the external momenta, we will calculate $g_{\rm{intra}}^{pp}(\textbf{q},\textbf{q}')-g_{\rm{intra}}^{pp}(p=2q_F)$ up to the leading order in the small parameters $\epsilon$ and $\cos\phi$. By keeping the small parameters in the upper and lower limits $\bar{l}_{1/2}$ while dropping those in the slowly varying logarithmic integrand, we obtain
\begin{align}
g_{\rm{intra}}^{pp}(\textbf{q},\textbf{q}')-g_{\rm{intra}}^{pp}(p=2q_F)
\sim-\frac{m^2U^3}{4\pi}\frac{\sqrt{4q_F^2-p^2}}{2q_F}
\label{eq:gintrappfinal}
\end{align}
for the regime of external momenta satisfying $\epsilon\ll 1$. 

The two-loop contributions to intra-pocket interaction involving only particle-hole bubbles, i.e. diagram (3i) and (3j), can be calculated in a similar way. In the same regime where the external momenta satisfy $\epsilon\ll1$, we obtain
\begin{align}
g_{\rm{intra}}^{ph}(\textbf{q},\textbf{q}')&=g_{3i}(\textbf{q},\textbf{q}')+g_{3j}(\textbf{q},\textbf{q}')\nonumber\\
&\propto-\frac{U^3m^2}{64\pi^3} (1-\frac{p^2}{4q_F^2}) \log [1-\frac{p^2}{4q_F^2}],
\label{eq:gintraphfinal}
\end{align}
where the minus sign is due to the closed fermion loop.

We then turn to the inter-pocket interaction. 
Among all the third-order contributions to $g_{\rm{inter}}^{(0)}$, diagram (3e) and (3f) in Supplementary Fig. \ref{diagram} are both just the product of two second-order corrections and do not contribute to momentum-dependence. 
Thus, we will focus only on diagram (3a)$\sim$ (3d). 
Note that similar to the case of intra-pocket interaction, diagram (3a) and (3b) involve vertex corrections from one particle-particle and one particle-hole bubble just like (3g) and (3h), while diagram (3c) and (3d) involve corrections from two particle-hole bubbles just like (3i) and (3j).
Thus, diagram (3a) and (3b) have similar amplitudes as diagram (3c) and (3d) except the momentum-transfer in the particle-hole bubble and the absence of closed fermion loop:
\begin{align}
g_{\rm{inter}}^{pp}(\textbf{q},\textbf{q}')-g_{\rm{inter}}^{pp}(p=2q_F)\sim\frac{m^2U^3}{4\pi}\frac{\sqrt{4q_F^2-p'^2}}{2q_F},
\label{eq:ginterpp}
\end{align}
where $\textbf{p}'\equiv\textbf{q}'+\textbf{q}$.
On the other hand, diagram (3c) and (3d) have the same amplitudes as diagram (3i) and (3j) such that
\begin{align}
g_{\rm{inter}}^{ph}(\textbf{q},\textbf{q}')&=g_{3c}(\textbf{q},\textbf{q}')+g_{3d}(\textbf{q},\textbf{q}')\propto-\frac{U^3m^2}{64\pi^3}(1-\frac{p^2}{4q_F^2}) \log [1-\frac{p^2}{4q_F^2}]
\label{eq:ginterph}
\end{align}
in the regime where $\epsilon$ is small. 
After collecting all the contributions, the $U^3$ corrections to the effective inter- and intra-pocket interactions at $E=\Lambda_0$ read 
\begin{align}
g_{\rm{inter}}^{(0),3}(\textbf{q},\textbf{q}')&
=g_{\rm{inter}}^{pp}(\textbf{q},\textbf{q}')+g_{\rm{inter}}^{ph}(\textbf{q},\textbf{q}')\nonumber\\
&\sim\frac{m^2U^3}{2\pi^3}\frac{\sqrt{4q_F^2-p'^2}}{2q_F}-\frac{U^3m^2}{64\pi^3}(1-\frac{p^2}{4q_F^2}) \log [1-\frac{p^2}{4q_F^2}]
\label{eq:ginter3}
\end{align}
and 
\begin{align}
g_{\rm{intra}}^{(0),3}(\textbf{q},\textbf{q}')&
=g_{\rm{intra}}^{pp}(\textbf{q},\textbf{q}')+g_{\rm{intra}}^{ph}(\textbf{q},\textbf{q}')\nonumber\\
&\sim-\frac{m^2U^3}{2\pi^3}\frac{\sqrt{4q_F^2-p^2}}{2q_F}-\frac{U^3m^2}{64\pi^3}(1-\frac{p^2}{4q_F^2}) \log [1-\frac{p^2}{4q_F^2}].
\label{eq:gintra3}
\end{align}

In summary, we have derived the effective inter- and intra-pocket interactions at $E=\Lambda_0$ from the bare repulsion $U>0$ up to two-loop order: 
\begin{align}
g_{\rm{inter}}^{(0)}(\textbf{q},\textbf{q}')&=g_{\rm{inter}}^{(0),1}(\textbf{q},\textbf{q}')+g_{\rm{inter}}^{(0),2}(\textbf{q},\textbf{q}') +g_{\rm{inter}}^{(0),3}(\textbf{q},\textbf{q}')\nonumber\\
&\sim C+\frac{m^2U^3}{2\pi^3}\frac{\sqrt{4q_F^2-p'^2}}{2q_F}-\frac{U^3m^2}{64\pi^3}(1-\frac{p^2}{4q_F^2}) \log [1-\frac{p^2}{4q_F^2}],
\label{eq:ginterall}
\end{align}
and 
\begin{align}
g_{\rm{intra}}^{(0)}(\textbf{q},\textbf{q}')&=g_{\rm{intra}}^{(0),1}(\textbf{q},\textbf{q}')+g_{\rm{intra}}^{(0),2}(\textbf{q},\textbf{q}') +g_{\rm{intra}}^{(0),3}(\textbf{q},\textbf{q}')\nonumber\\
&\sim C'-\frac{m^2U^3}{2\pi^3}\frac{\sqrt{4q_F^2-p^2}}{2q_F}-\frac{U^3m^2}{64\pi^3}(1-\frac{p^2}{4q_F^2}) \log [1-\frac{p^2}{4q_F^2}],
\label{eq:gintraall}
\end{align}
where $C>0$ and $C'<0$ are momentum-independent constants.\\

{\bf\normalsize{Supplementary Note 3: The real-space profile of the phase of the intra-pocket pairing wavefunction}}\\
\begin{figure}[t]
\begin{centering}
\includegraphics[width=0.9\linewidth]{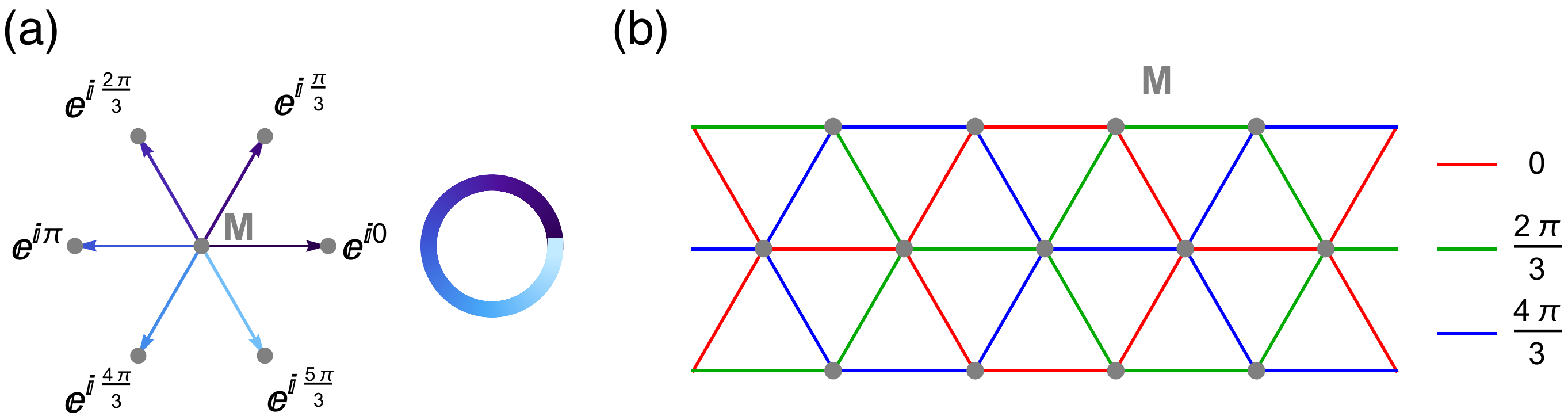}
\end{centering}
\caption{The phase of the intra-pocket pairing wave function in real space: (a) shows the phase $e^{i\theta_{\textbf{d}}}$ in Supplementary Eq. \eqref{eq:intrawfrealspace} which accounts for the chiral $\tilde{l}=1$ phase-winding within a pocket. The arrows represent $\textbf{d}$ for the nearest-neightboring transition metal ions $\rm{M}$ (the grey dots). (b) shows the spatially modulated phase $e^{i2\textbf{K}\cdot\textbf{r}}$ due to the finite pair-momentum $2\textbf{K}$ for a spin-up pair. We consider only pairing between electrons from nearest-neightboring sites. Thus, the phase of the pairing wave function is defined on each bond. The colors on the bonds represent different values of $2\textbf{K}\cdot\textbf{r}$.}
\label{phase}
\end{figure}
Since the intra-pocket pairs are spinless and the intra-pocket interaction is attractive in the $\tilde{l}=1$ channel, the intra-pocket pairing wavefunction on the spin-up pocket is expected to be $\Delta^{\uparrow\uparrow}_{\textbf{q}}=\langle \psi_{\textbf{K}+\textbf{q},\uparrow}\psi_{\textbf{K}-\textbf{q},\uparrow}\rangle\propto q_x\pm iq_y$ in terms of separate spin and valley indices. The p-wave pairing is expected to be chiral to avoid nodes due to energetics. 
The pairing wavefunction in real space can then be obtained by doing the following Fourier transform:
\begin{align}
\langle \psi_{\textbf{r}+\frac{d}{2},\uparrow}\psi_{\textbf{r}-\frac{d}{2},\uparrow}\rangle=\sum_{\textbf{q}}\langle \psi_{\textbf{K}+\textbf{q},\uparrow}\psi_{\textbf{K}-\textbf{q},\uparrow}\rangle e^{i2\textbf{K}\cdot\textbf{r}}e^{i\textbf{q}\cdot\textbf{d}}
=e^{i2\textbf{K}\cdot\textbf{r}}\sum_{\theta_{\textbf{q}}}q_Fe^{\pm i\theta_{\textbf{q}}}e^{iqd\cos(\theta_{\textbf{q}}-\theta_{\textbf{d}})}
\propto e^{i2\textbf{K}\cdot\textbf{r}}e^{i\theta_{\textbf{d}}}
\label{eq:intrawfrealspace}
\end{align} 
where $\textbf{r}$ and $\textbf{d}=d(\cos\theta_{\textbf{d}},\sin\theta_{\textbf{d}})$ are the center-of-mass and relative positions of the pair repectively, the relative momentum $\textbf{q}=q_F(\cos\theta_{\textbf{q}},\sin\theta_{\textbf{q}})$ is confined on the circular pocket centered at $\textbf{K}$ with $q_F$ being the Fermi momentum, and $\theta_{\textbf{d}}$($\theta_{\textbf{q}}$) is the angle between $\textbf{d}$($\textbf{q}$) and $\textbf{K}$.
While the phase winding from $e^{i\theta_{\textbf{d}}}$ [see Supplementary Fig. \ref{phase}(a)] accounts for the $q_x+iq_y$ pairing symmetry on a pocket, the spatial modulation in phase from $e^{i2\textbf{K}\cdot\textbf{r}}$ [see Supplementary Fig. \ref{phase}(b)] is a consequence of the finite pair-momentum $2\textbf{K}$. 


\end{document}